\documentclass[12pt,preprint]{aastex}

\usepackage{amssymb}
\usepackage{amsmath}
\usepackage{graphicx}
\usepackage{subfigure}
\usepackage{latexsym}
\usepackage{xcolor}
\usepackage{float}

\shorttitle{Kelvin-Helmholtz instability in rotating solar jets}
\shortauthors{Zaqarashvili et al.}

\begin{document}

\title{Rieger-type periodicity during solar cycles 14-24: estimation of dynamo magnetic field strength in the solar interior
}
\author{Eka Gurgenashvili\altaffilmark{1}, Teimuraz V. Zaqarashvili\altaffilmark{2,1,5}, Vasil Kukhianidze\altaffilmark{1}, Ramon Oliver\altaffilmark{3,4}, Jose Luis Ballester\altaffilmark{3,4}, Giorgi Ramishvili\altaffilmark{1}, Bidzina Shergelashvili\altaffilmark{5,1,6}, Arnold Hanslmeier\altaffilmark{2}, and Stefaan Poedts\altaffilmark{6}}

\altaffiltext{1}{Abastumani Astrophysical Observatory at Ilia State University, Tbilisi, Georgia}
\altaffiltext{2}{IGAM, Institute of Physics, University of Graz, Universit{\"a}tsplatz 5, 8010 Graz, Austria, Email: teimuraz.zaqarashvili@uni-graz.at}
\altaffiltext{3}{Departament de F\'isica, Universitat de les Illes Balears, E-07122, Palma de Mallorca, Spain}
\altaffiltext{4}{Institute of Applied Computing and Community Code (IAC$^3$), UIB, Spain}
\altaffiltext{5}{Space Research Institute, Austrian Academy of Sciences, Schmiedlstrasse 6, 8042 Graz, Austria}
\altaffiltext{6}{Centre for Mathematical Plasma Astrophysics, Department of Mathematics, KU Leuven, Celestijnenlaan 200B, 3001, Leuven, Belgium}

\begin{abstract}

Solar activity undergoes a variation over time scales of several months known as Rieger-type periodicity, which usually occurs near maxima of sunspot cycles. An early analysis showed that the periodicity appears only in some cycles, and is absent in other cycles. But the appearance/absence during different cycles has not been explained. We performed a wavelet analysis of sunspot data from the Greenwich Royal Observatory and the Royal Observatory of Belgium during cycles 14-24. We found that the Rieger-type periods occur in all cycles, but they are cycle-dependent: shorter periods occur during stronger cycles. Our analysis revealed a periodicity of 185-195 days during the weak cycles 14-15 and 24, and a periodicity of 155-165 days during the stronger cycles 16-23. We derived the dispersion relation of the spherical harmonics of the magnetic Rossby waves in the presence of differential rotation and a toroidal magnetic field in the dynamo layer near the base of the convection zone. This showed that the harmonic of fast Rossby waves with m=1 and n=4, where m (n) indicate the toroidal (poloidal) wavenumbers, respectively, perfectly fit with the observed periodicity.
The variation of the toroidal field strength from weaker to stronger cycles may lead to the different periods found in those cycles, which explains the observed enigmatic feature of the Rieger-type periodicity. Finally, we used the observed periodicity to estimate the dynamo field strength during cycles 14-24. Our estimations suggest a field strength of 40 kG for the stronger cycles, and 20 kG for the weaker cycles.

\end{abstract}

\keywords{Sun: activity --- Sun: interior --- Sun: oscillations}

\section{Introduction}\label{intro}

The solar magnetic activity, which is manifested by the emergence of sunspots and active regions and the consequent energy release by flares, undergoes variations over many different time scales. The main periodicity in number of sunspots occurs on a time scale of 11 years, which is called the solar cycle or Schwabe cycle \citep{Schwabe1844}. The other time scales that occur, can be divided into longer and shorter time scales. The longer time scales are of the order of hundreds and even thousands of years \citep{Gleissberg1939,Suess1980,Solanki2004,Usoskin2004,Hanslmeier2013,zaqarashvili2015}, while the shorter time scales that occur are of the order of months (155-160 days) known as the Rieger periodicity \citep{Rieger1984,Lean1989,Carbonell1990,Oliver1998}. Another variation of the solar activity occurs over $\sim$ 2 years, which is known as the quasi-biennial oscillation \citep{sakurai81,gigolashvili95,vecchio09,zaqarashvili2010b,McIntosh2015}. The short-period variation may appear as a clue to understand the basics of solar activity \citep{McIntosh2015}. Therefore, these variations need to be studied in detail. In the remaining part of the present paper, we focus on the Rieger-type periodicity.

The periodicity of 154 days was first found by \citet{Rieger1984} in the $\gamma$-ray flares observed by the Solar Maximum Mission (SMM) near the maximum of solar cycle 21. The analysis of different data sets during the past few cycles confirmed the existence of Rieger-type periods in different indicators of solar magnetic activity, viz.\ in X-ray flares \citep{Dennis1985,Bai1987,Kile1991,Dimitropolou2008}, in the occurrence rates of solar flare energetic electrons \citep{Droge1990}, in sunspot group numbers \citep{Lean1989,Carbonell1990,Lean1990,Carbonell1992,Oliver1998,Ballester1999}, in type II and IV radio bursts \citep{Verma1991}, in type III radio bursts \citep{Lobzin2012}, and in microwave \citep{Kile1991} and proton \citep{Bai1990} flares. \citet{Bogart1985} investigated the occurrence of flares inferred from microwave data and found a strong confirmation of a 152  day periodicity. \citet{Bai1987} reported an analysis of major flares observed with the Hard X-ray Burst Spectrometer (HXRBS) aboard SMM and conclude that the 152-day periodicity is a global phenomenon. Therefore, the underlying cause of this periodicity must be a mechanism involving the whole Sun. Rieger-type periods of 152-158 days were found in solar data which are related to strong magnetic fields on the surface \citep{Bouwer1992,Pap1990,Verma1991,Oliver1998,Ballester1999,Krivova2002}. \citet{Lean1989} found that the periodicity was present in the sunspot blocking function, in the 10.7 cm radio flux and in sunspot numbers, while it was not significant in the case of the plage index. This suggests that Rieger periodicity is not only a feature of flare activity, but that it is also associated with regions of compact magnetic field structures. Therefore, the periodicity should be connected to the location where the strong magnetic field is generated. In a series of papers, \citet{Delache1985,Delache1988a,Delache1988b} detected the Rieger periodicity in measurements of the solar diameter variations, the total irradiance, the 10.7 cm radio flux and the 10830 ${\AA}$ Helium line. \citet{Chowdhury2013} studied  time series of the full disk integrated soft and hard X-ray emission from the solar corona during the descending phase of solar cycle 23. \citet{Chowdhury2015} conducted Lomb-Scargle periodogram and wavelet analyses of the sunspot area/number, the 10.7 cm solar radio flux, the average photospheric magnetic flux, the interplanetary magnetic field and the geomagnetic activity index during the ascending phase of the current solar cycle 24, and found a Rieger periodicity of 150-160 days and a near Rieger periodicity of 130-190 days.

Many authors have studied the long-term appearance of the Rieger-type periodicity. \citet{Lean1990} examined sunspot area data during the solar cycles 12-21 and found that the periodicity usually appears near the cycle maxima. It was shown that the 155 day periodicity occurs during intervals of 1 to 3 years, and that it may vary from 130 to 185 days. A similar result was found by \citet{Oliver1998} and \citet{zaqarashvili2010a} through a wavelet analysis of sunspot area. \citet{Carbonell1990} and \citet{Carbonell1992} presented evidence of the 155 days periodicity in records of the sunspot area during the cycles 14 to 20 and 12 to 21, respectively. They found that the Rieger periodicity was clearly seen during cycles 16-21, while it was absent during the cycles 12-15. \citet{Ballester2002} analyzed the historical records of photospheric magnetic flux and found that the periodicity appeared during cycle 21, but that it was absent in  cycle 22. \citet{Akimov2012} used a wavelet and Fourier analysis to investigate the temporal behavior of the 156 day periodicity for series of daily solar indices: the Wolf numbers for 161 years (from 1849), the solar flux at $2800\;$MHz F10.7 for 63 years (from 1947), the number of X-ray flares for 29 years (from 1981), and the number of optical flares for 11 years in cycle 21. Hence, all these previous studies showed that the Rieger periodicity of 154 day is not a permanent feature of the solar activity, but that it varies from cycle to cycle.

The physical reason for the occurrence of the Rieger-type periodicity is not completely clear. Several different mechanisms have been suggested to explain the enigmatic features of the periodicity. \citet{Ichimoto1985} suggested that the 155-day periodicity may be connected to the timescale of the storage and/or the escape of the magnetic fields in the solar convection zone. \citet{Bai1991} supposed a clock modelled by an oblique rotator or oscillator with a period of 25.8 days, and suggested that the periodicity of 155-160 days is just a subharmonic of that fundamental period. \citet{Wolff1992} and \citet{Sturrock2013,Sturrock2015} suggested that the periodicity can be explained in terms of r-mode oscillations (which are actually hydrodynamic (HD) Rossby or planetary waves) of the solar interior. \citet{Lou2000} suggested that the Rieger periodicity can be related to large-scale equatorially trapped HD Rossby-type waves in the solar photosphere. However, the HD Rossby wave theory does not take into account the magnetic field, therefore it is unclear how the waves can modulate the solar magnetic activity. Recently, \citet{zaqarashvili2010a} suggested that the Rieger-type periods of 155-160 days could be explained by magnetic Rossby waves in the solar tachocline, which are unstable owing to the differential rotation and the toroidal magnetic field. Unstable harmonics of magnetic Rossby waves lead to the periodic emergence of magnetic flux at the solar surface due to the magnetic buoyancy, which causes the observed periodicity in the magnetic activity. The dispersion relation of the magnetic Rossby waves depends on the unperturbed magnetic field strength \citep{zaqarashvili2007,zaqarashvili2009}. Therefore, the possible variation of the mean dynamo magnetic field from cycle to cycle may influence the Rieger periodicity, which could be correlated with the strength of the solar cycle. The observed periodicity can then be used to estimate the magnetic field strength in the solar dynamo layer near the base of the convection zone, using the dispersion relation of the magnetic Rossby waves. The strength of the dynamo magnetic field is crucial to know which dynamo model works in the solar interior \citep{Fan2009,Charbonneau2010,Charbonneau2013}. Therefore, probing the dynamo layer by exploiting the Rieger periodicity may appear as a very important tool to better understand the solar magnetic activity.

In this paper, we first analyze the sunspot data of the Greenwich Royal Observatory (GRO) and Royal Observatory of Belgium (ROB) for cycles 14-24, using the Morlet wavelet tool \citep{Torrence1998}, in order to find Rieger-type periods during the different cycles and to establish their relation with the cycle strength. Then, we derive the dispersion relation of the magnetic Rossby waves in the presence of a toroidal magnetic field and differential rotation. Finally, we estimate the strength of the toroidal magnetic field in the solar dynamo layer during the cycles 14-24, using the dispersion relation of the fast magnetic Rossby waves.

\section{Main equations}

In order to study the dependence of the Rieger-type periodicity on the solar cycle strength, we use two data sets, viz.\ 1)~the Greenwich Royal Observatory (GRO) USAF/NOAA sunspot data (\url{http://solarscience.msfc.nasa.gov/greenwch.shtml}); and 2)~the  WDC-SILSO, Royal Observatory of Belgium (ROB) new sunspot number data (\url{http://www.sidc.be/silso/datafiles}). The GRO data sets starts from 1874 (which corresponds to cycle 12) and runs till 2015, but we here use only data from July 1901 till December 2015 (which corresponds to the period from the minimum of cycle 14 till the descending phase of cycle 24) because of the data gaps that occur in cycles 12-13. The ROB data starts from 1818 and runs till 2015, and covers the cycles 10-24. However, we here only use the same interval from 1900 to 2015, in order to compare with the GRO data.

We use a Morlet wavelet analysis to search for the Rieger periodicity in the solar cycles 14-24 in both data sets. Figure~1 shows the power spectra of the sunspot area variations through the different solar cycles using the GRO (middle panel) and the ROB (lower panel) daily data. In this figure, it is clearly seen that oscillations with Rieger-type periods are localized near the cycle maxima, as was previously discovered \citep{Lean1990,Oliver1998,zaqarashvili2010a}. The periodicity changes from cycle to cycle and one can easily see (even with the naked eye) that the stronger cycles generally support oscillations with shorter periods. For example, cycle 21, during which the Rieger-type periodicity was discovered \citep{Rieger1984}, was one of the strongest cycles in the 20th century and a periodicity of 155-160 days has been found in both data sets. A similar periodicity appears during the other strong cycles 17 and 18. On the other hand, the weaker cycles (14, 15, and 24) show oscillations with a longer periodicity. Therefore, the Rieger periodicity is clearly correlated with the solar cycle strength. To estimate the Rieger periods during each cycle, we performed a Morlet wavelet analysis in each cycle separately of both the GRO and ROB data. The Rieger periods that resulted from the global wavelet spectra during cycles 14-24, which cover an interval of 150-200 days, are given in Table~1 and shown on Fig.~2. Both the GRO and ROB data display similar periods in each cycle (with a difference of only a few days). We note that some oscillatory power is located in periodicities longer than 200 days (it is more clearly seen on upper panel of Figure~1), which correspond to the longer period branch like the quasi-annual oscillations found by \citet{McIntosh2015}. The goal of this paper, however, is to study the Rieger periodicity only. Therefore, we did not show the upper part of the power spectrum ($>$ 200 days), which has a complex nature and may have a different physical origin. We will study the complete spectrum of oscillations in the near future.

\begin{figure}
\includegraphics[width = 4.5in]{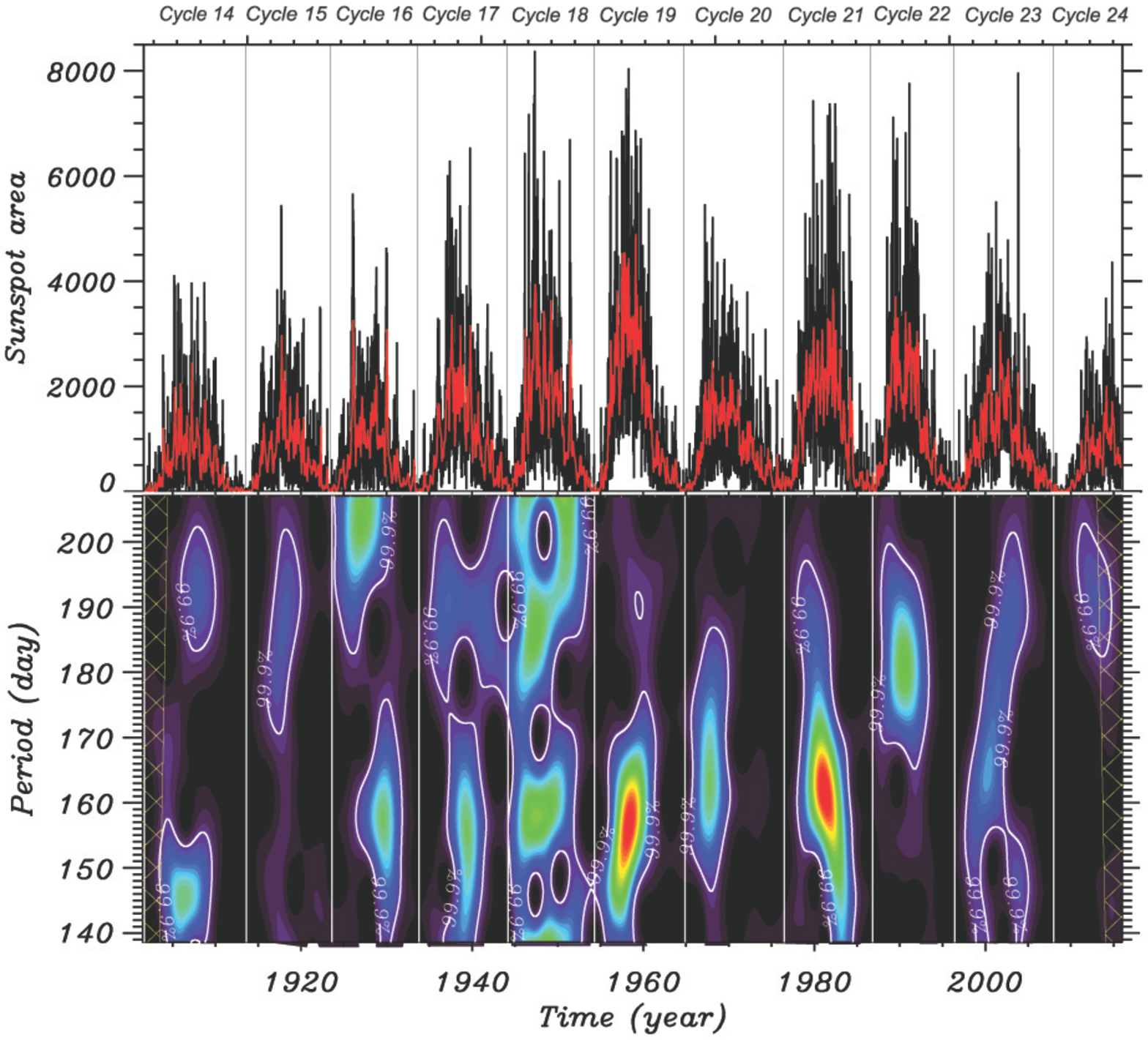}
\includegraphics[width = 4.5in]{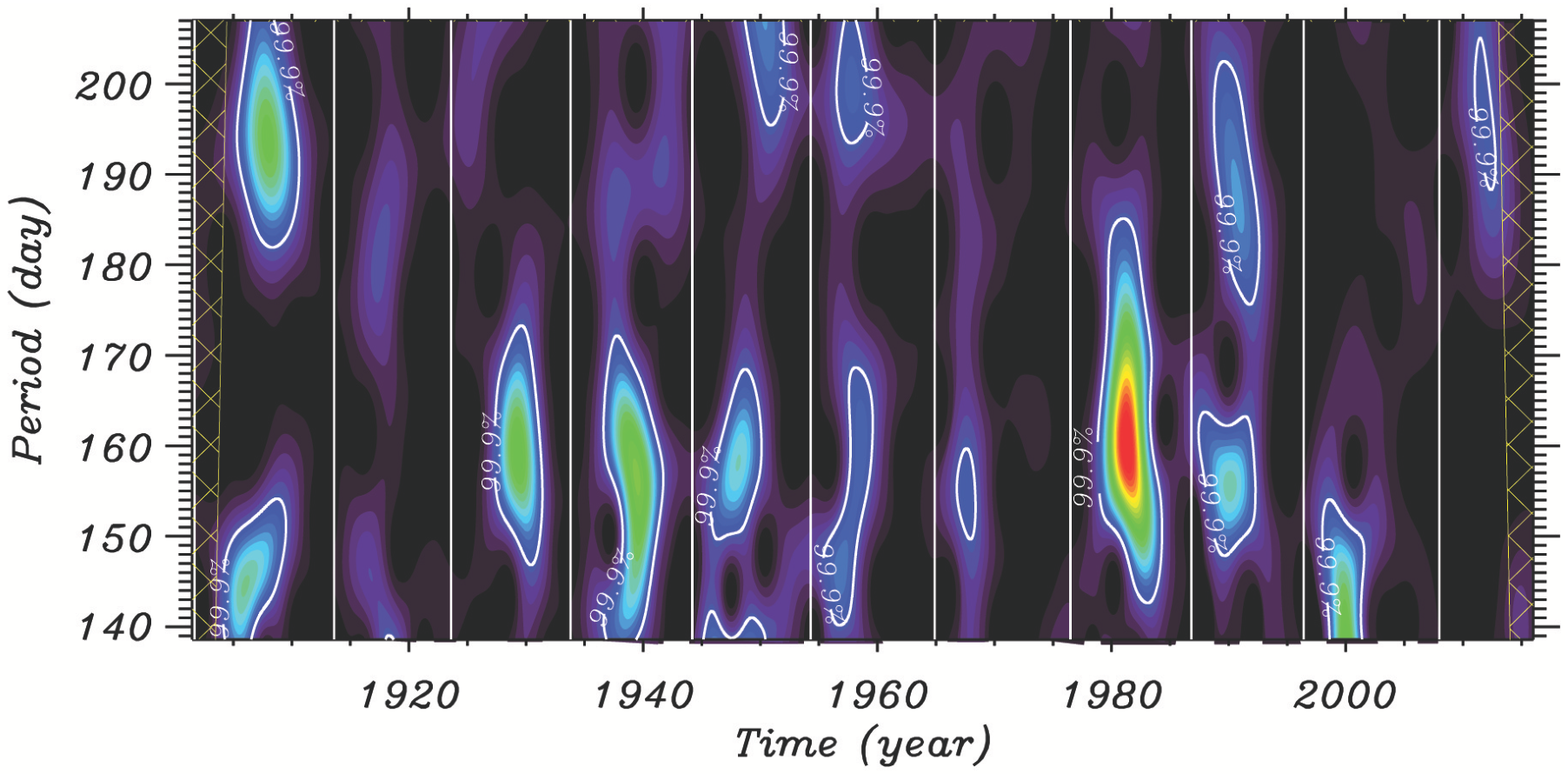}
\caption{Rieger periodicity during solar cycles 14-24. Top panel: daily and monthly averaged sunspot area data with black and red color, respectively, from GRO data. Middle panel: Morlet wavelet analysis of GRO daily sunspot data with $k_0=23$ \citep{Torrence1998}. Vertical solid lines correspond to solar activity minimum. White lines encircle the most important powers, above confidence level 99.9\%. Lower panel: Morlet wavelet analysis of ROB daily sunspot data. Hatched areas outside the cone of influence (COI) indicate the regions where wavelet transform is not reliable. }
\end{figure}

Figure~2 clearly shows that the Rieger periodicity is correlated with the long-term variations in the solar cycle strength. The solar cycle itself is modulated by the so-called Gleissberg period of about 100 years \citep{Gleissberg1939,Hathaway2010,zaqarashvili2015}, which is also seen in long term evolution of sunspot data (e.g.\ upper panel of Fig.~1). The Rieger periodicity displays a similar long-term evolution: a longer periodicity corresponds to weaker cycles. The correlation coefficient between the Rieger periodicity from the GRO (ROB) data and the sunspot number is estimated as -0.64 (-0.67), respectively, which is a rather strong correlation \citep{Cohen1988}. This correlation probably indicates a connection of the periodicity to the region in the solar interior, where the dynamo magnetic field is generated. \citet{zaqarashvili2010a} explained the occurrence of the Rieger periodicity in terms of Rossby waves in the tachocline, which are influenced by the large-scale toroidal dynamo magnetic field. Then, the variation of the mean toroidal magnetic field strength through different cycles can be responsible for the observed different periods in sunspot data. The dispersion relation of spherical magnetic Rossby waves was derived by \citet{zaqarashvili2007} in the simplest case of rigid rotation and a homogeneous magnetic field. Then, \citet{zaqarashvili2009} were able to derive the dispersion relation in the case of a more realistic latitudinal profile of the toroidal magnetic field, which resembles the dynamo magnetic field. In both cases, the dispersion relations were obtained for a rigid rotation. However, observations show that the solar rotation has a latitudinal profile being faster at the equator and slower at poles. The differential rotation is, however, solar-cycle dependent: the equatorial rotation and the parameters of the differential rotation vary from cycle to cycle
\citep{Javaraiah2005,Brajsa2006}. The differential rotation may influence the spectrum of magnetic Rossby waves and, hence, should be taken into account when deriving the dispersion relation.

\begin{figure}
\includegraphics[width=\columnwidth]{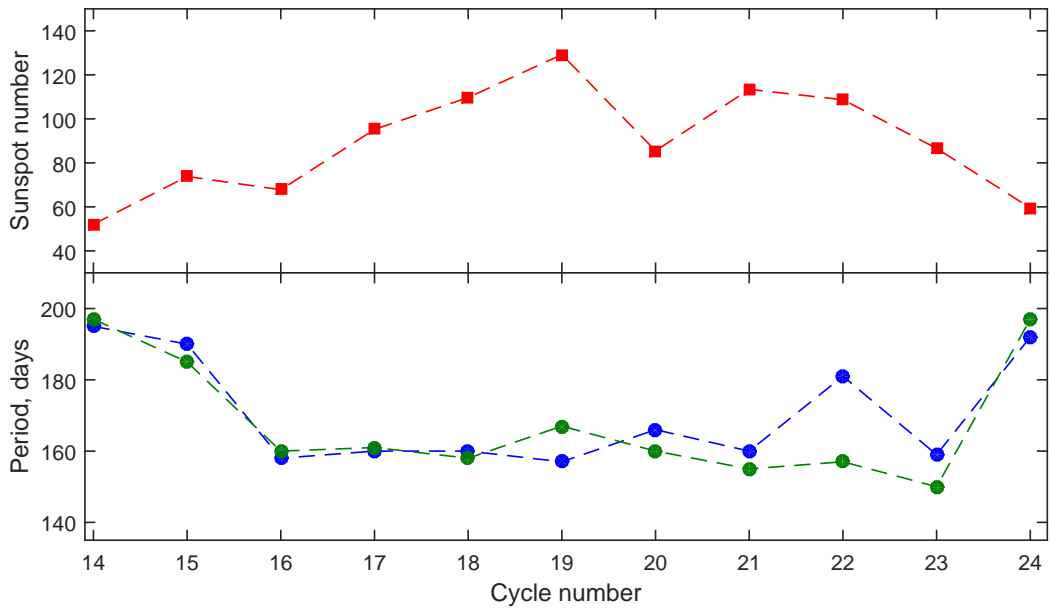}
\caption{Rieger periods from global wavelet spectra during solar cycles 14-24. Blue circles (green circles) denote Rieger periods found in GRO (ROB) data in the corresponding cycle. Red squares correspond to the sunspot number obtained after averaging of each cycles from ROB data. }
\end{figure}

The observed profile of the solar differential rotation is given by
\begin{equation}
\label{dif} \Omega=\Omega_0 (1-s_2 \cos^2\theta-s_4\cos^4\theta),
\end{equation}
where $\theta$ denotes the co-latitude, $\Omega_0$ is the equatorial angular velocity, while $s_2$ and $s_4$ are constant parameters. The last term in the expression above is especially important at high latitudes, but has less influence at lower latitudes. Moreover, the observation of the differential rotation at higher latitudes is uncertain to some extent which may cause  significant errors in the estimation of the parameter $s_4$. Therefore, many authors neglect this term when they construct the differential rotation profile. For simplicity of the further calculations, we here also neglect the term $-s_4\cos^4\theta$, and hence the following simplified differential rotation profile is used throughout the remainder of the present paper
\begin{equation}
\label{dif1} \Omega=\Omega_0 (1-s_2 \cos^2\theta).
\end{equation}
This expression is  more justified  at lower latitudes, where the main magnetic activity takes place.

In the next section, we derive the dispersion relation for the magnetic Rossby waves including the effect of differential rotation.
\section{Dispersion relation for magnetic Rossby waves in the presence of differential rotation}

Rossby waves (or planetary waves) are large-scale waves, that occur due to the latitudinal variation of the Coriolis parameter on rotating spheres \citep{Rossby1939}. The waves determine the large-scale dynamics of the Earth atmosphere and oceans \citep{Gill1982}. The inclusion of the large-scale toroidal magnetic field influences the dynamics of the Rossby waves in plasma, leading to the splitting of the hydrodynamic mode into fast and slow magnetic Rossby modes \citep{zaqarashvili2007}. Therefore, the strength of the magnetic field generally determines the frequency spectrum of the magnetic Rossby waves. We use the spherical coordinate system $(r,\theta,\varphi)$, where $r$ denotes the distance from the solar center, $\theta$ is the co-latitude and $\varphi$ is the longitude. We assume that the unperturbed magnetic field has only a toroidal component
\begin{equation}
\label{Btheta} B_{\varphi}(\theta) ={\tilde B_{\varphi}}(\theta) \sin \theta.
\end{equation}
The dispersion relation of the Rossby waves can be derived from the shallow water MHD equations, which include the radial coordinate as the height of a shallow layer. The consideration of a shallow layer leads to the appearance of surface gravity or Poincar\'e waves in the dispersion relation. For a negligible thickness of the layer, however, the surface gravity waves disappear and only the Rossby waves remain \citep{zaqarashvili2009}. Hence, the consideration of a spherical surface $\theta,\varphi$ correctly describes the main properties of Rossby waves for a negligible thickness of the layer with regards to the solar radius. Therefore, we use the 2D incompressible linearized magnetohydrodynamic (MHD) equations in the frame rotating with $\Omega_0$ \citep{zaqarashvili2010a}
\begin{equation}
\label{eq1} {{\partial u _\theta}\over {\partial t}}+{\Omega_1(\theta)}{{\partial u _\theta}\over {\partial \varphi}}-{2\Omega\cos \theta u_\varphi}=\\
  {-{{1\over {\rho R_0}}{{\partial p_T} \over {\partial \theta}}+ {{{\tilde B_{\varphi}}(\theta)} \over {4 \pi \rho R_0}} {{\partial b_\theta} \over {\partial \varphi}}-2 {{{\tilde B_{\varphi}}(\theta)\cos\theta} \over {4\pi \rho R_0}}b_\varphi}},
\end{equation}
\begin{multline}
\label{eq2}
{{\partial u_\varphi} \over {\partial t}}+ {\Omega_1 ({\theta})} {{\partial u _\varphi}\over {\partial \varphi}} + {2\Omega_0\cos \theta u_\theta}+ {\Omega_1 ({\theta})\cos \theta u_\theta} + {u_\theta}{{\partial}\over {\partial \theta}} { [\sin\theta\Omega_1({\theta})]} =
 -{1\over {\rho R_0 \sin\theta}}{{\partial p_T} \over {\partial \varphi}} \\
   + {{{\tilde B_{\varphi}}(\theta)} \over {4 \pi \rho R_0}} {{\partial b_\varphi} \over {\partial \varphi}}+ {{b_\theta} \over {4 \pi \rho R_0 \sin\theta}} {{\partial } \over {\partial \theta}} {[{{\tilde B_{\varphi}}(\theta)}\sin^2\theta]},
\end{multline}
\begin{equation}
\label{eq3}  {{\partial b_\theta} \over {\partial t}} + {\Omega_1 ({\theta})} {{\partial b _\theta}\over {\partial \varphi}}=  {{{{\tilde B_{\varphi}}(\theta)}} \over { R_0 }} {{\partial u_\theta} \over {\partial \varphi}},
\end{equation}
\begin{equation}
\label{eq4} {{\partial } \over {\partial \theta}} {[\sin\theta b_{\theta}]}+ {{\partial b_{\varphi}} \over {\partial \varphi}} =0,\,\,{{\partial } \over {\partial \theta}} {[\sin\theta u_{\theta}]}+ {{\partial u_{\varphi}} \over {\partial \varphi}} =0,
\end{equation}
where $u_{\theta}$, $u_{\varphi}$, $b_{\theta}$, $b_{\varphi}$ denote the velocity and magnetic field perturbations, $p_T$  is the perturbation of total (thermal + magnetic) pressure, $\rho$ is the density perturbation, $R_0$ denotes the distance from the center and $\Omega_1 {(\theta)}=-\Omega_0s_2 \cos^2\theta$. Note, that in the denominator of the first term at the right hand side of Eq.~(3) in \citet{zaqarashvili2010a} $\rho$ is missing, but this is merely typo and all the consequent equations in that paper are correct. Considering the stream functions for the velocity and the magnetic field,
\begin{equation}
\label{stream} u_\theta={1 \over {\sin\theta}} {{\partial \psi} \over {\partial \varphi}},\,\,u_\varphi= - {{\partial \psi} \over {\partial \theta}},\,\,b_\theta={1 \over {\sin\theta}} {{\partial \phi} \over {\partial \varphi}},\,\,b_\varphi= - {{\partial \phi} \over {\partial \theta}}
\end{equation}
and using a Fourier analysis of the form $exp[i(m\varphi-\omega t)]$,  we derive the two following equations \citep{zaqarashvili2010a}
\begin{equation}
\label{e1} B\psi=(\Omega_d-{\hat \omega})\phi,
\end{equation}
and
\begin{multline}
\label{e2} (\Omega_d-{\hat \omega})\left[{{\partial} \over {\partial \mu}} {(1-\mu^2)} {{\partial}\over{\partial\mu}} - {{m^2}\over{1-\mu^2}}\right]\psi + \left(2-{{d^2}\over{d\mu^2}} {[\Omega_d(1-\mu^2)]}\right)\psi \\
-\beta^2 B \left[ {{\partial} \over {\partial \mu}} {(1-\mu^2)} {{\partial}\over{\partial\mu}} - {{m^2}\over{1-\mu^2}}\right]\phi+ \beta^2{{d^2}\over{d\mu^2}} {[B(1-\mu^2)]}\phi=0,
\end{multline}
where $\mu=\cos\theta$, $\psi$ is normalized by $\Omega_0 R_0$, $\phi$ is normalized by $B_0$, and
\begin{equation}
\label{par} \Omega_d({\mu})={{\Omega_1({\mu})} \over {\Omega_0}},  {\hat \omega}={{\omega}\over{m\Omega_0}}, \beta^2={{B^2_0}\over {4\pi\rho\Omega^2_0 R^2_0}}, B(\mu)={{\tilde B_{\varphi}(\mu)}\over {B_0}}.
\end{equation}

Now we define a new function $H$ as
\begin{equation}
\label{H}
\psi=(\Omega_d-{\hat \omega})H,\,\,\phi=BH.
\end{equation}
Equations~(\ref{e1})-(\ref{e2}) can then be cast into one equation, viz.\
\begin{multline}
\label{eq-f} {{\partial}\over{\partial\mu}} {(1-\mu^2)}{{\partial H}\over {\partial\mu}} +{P^{\prime}\over P}(1-\mu^2){{\partial H}\over {\partial\mu}} +\left [-{{m^2}\over{1-\mu^2}}+{{2(\Omega_d-{\hat \omega})[1+(\mu\Omega_d)^{\prime}]-2\beta^2B(\mu B)^{\prime}}\over P}\right ]H=0,
\end{multline}
where $P(\mu)=(\Omega_d-{\hat \omega})^2-\beta^2B^2$ and $^{\prime}$ means differentiation by $\mu$.

Substitution of a new function
\begin{equation}
\label{H}
H={\tilde H}(\mu)\exp\left [-{1\over 2}\int{{{P^{\prime}}\over P}d\mu}\right ]
\end{equation}
gives the equation
\begin{multline}
\label{eq-fin} {{\partial}\over{\partial\mu}} {(1-\mu^2)}{{\partial {\tilde H}}\over {\partial\mu}}+ \\ \left [-
{{m^2}\over{1-\mu^2}}+{1\over 4}{{P^{\prime 2}}\over {P^2}}(1-\mu^2)+ {{2\mu P^{\prime}-P^{\prime \prime}(1-\mu^2)+4(\Omega_d-{\hat \omega})[1+(\mu\Omega_d)^{\prime}]-4\beta^2B(\mu B)^{\prime}}\over {2P}}\right ]{\tilde H}=0.
\end{multline}

At this stage, we fix the toroidal magnetic field configuration using the profile suggested by \citet{Gilman1997}
\begin{equation}
\label{B}
B_{\varphi}={\tilde B_{\varphi}}\sqrt{1-\mu^2}=B_0 \mu \sqrt{1-\mu^2},
\end{equation}
which has  maxima at mid latitudes, becomes zero at the poles and changes sign at the equator. Therefore, it generally resembles the solar dynamo magnetic field.

In this case, Eq.~(15) can be replaced by
\begin{multline}
\label{eq-B1} {{\partial}\over{\partial\mu}} {(1-\mu^2)}{{\partial {\tilde H}}\over {\partial\mu}}+ [-
{{m^2}\over{1-\mu^2}}+ \left ({{2s_2\mu({\hat \omega}+s_2 \mu^2)-\beta^2 \mu}\over {({\hat \omega}+s_2 \mu^2)^2-\beta^2\mu^2}}\right )^2(1-\mu^2)+\\ {{-2(1+s_2){\hat \omega}-(2s_2-12s_2{\hat \omega}+6s^2_2+5\beta^2)\mu^2+16s^2_2\mu^4+\beta^2}\over {({\hat \omega}+s_2 \mu^2)^2-\beta^2\mu^2}}]{\tilde H}=0.
\end{multline}

\begin{figure}
\includegraphics[width=\columnwidth]{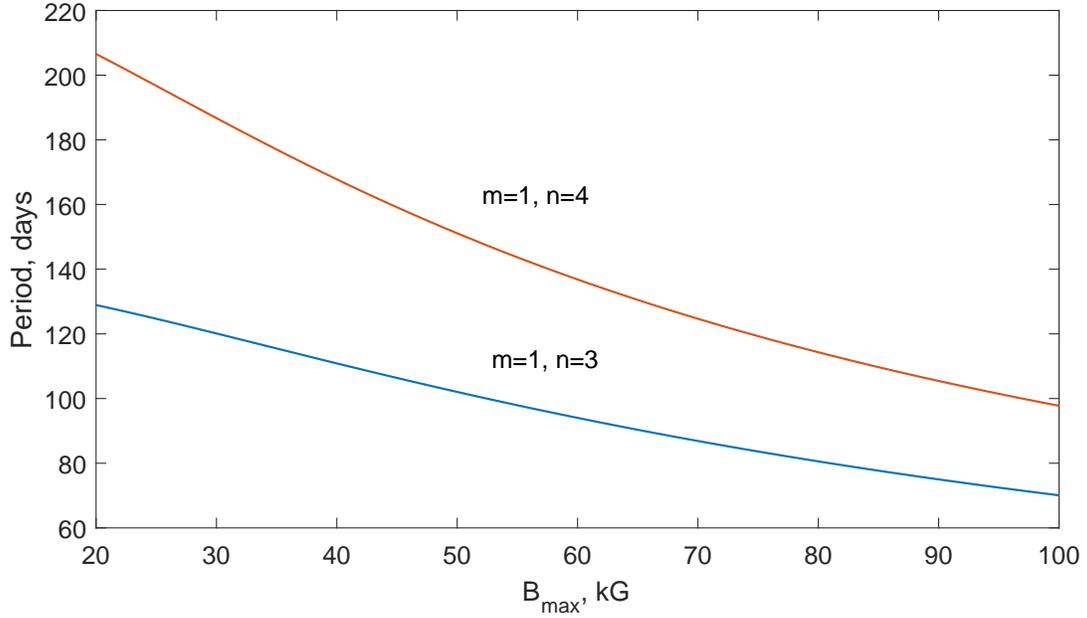}
\caption{Period of fast magnetic Rossby waves vs toroidal magnetic field strength, $B_{max}$. Red (blue) line shows the period of $m=1$ and $n=4$ ($n=3$) spherical harmonic. Here we use the following parameters typical for the base of convection zone: $\rho=$ 0.2 g cm$^{-3}$, $R_0=5$ 10$^{10}$ cm, $\Omega_0=2.7$ 10$^{-6}$ s$^{-1}$ and $s_2=0.17$.}
\end{figure}

For weak differential rotation $s_2 << 1$, and low latitudes $\mu << 1$, this equation is transformed into the following equation
\begin{equation}
\label{eq-B11} {{\partial}\over{\partial\mu}} {(1-\mu^2)} {{\partial{\tilde H}}\over {\partial\mu}}+\left[ -{{m^2}\over{1-\mu^2}}-{{2(1+s_2){\hat \omega}-\beta^2}\over{{\hat \omega}^2}}\right]{\tilde H}=0.
\end{equation}
This equation is an associated Legendre equation when
\begin{equation}
\label{eq-B12} - {{2(1+s_2){\hat \omega}-\beta^2}\over{{\hat \omega}^2}}=n(n+1)
\end{equation}
and its typical solutions are the associated Legendre functions $P^m_n(\mu), Q^m_n(\mu)$, where $m \geq 1$ and $n\geq 1$ are integers. Eq.~(19) defines the dispersion relation for fast and slow magnetic Rossby waves
\begin{equation}
\label{quadratic} n(n+1)\omega^2+2 m \Omega_0 {(1+s_2)\omega}-{{B^2_0 m^2} \over {4 \pi \rho  R^2_0 }}=0.
\end{equation}
This equation is similar to Eq.~(5) in \citet{zaqarashvili2009}, which was derived for a rigid rotation, with the only difference that the second term is multiplied by $1+s_2$. For the rigid rotation  $s_2=0$, so that the two equations are identical in that case.

\begin{figure}
\includegraphics[width=\columnwidth]{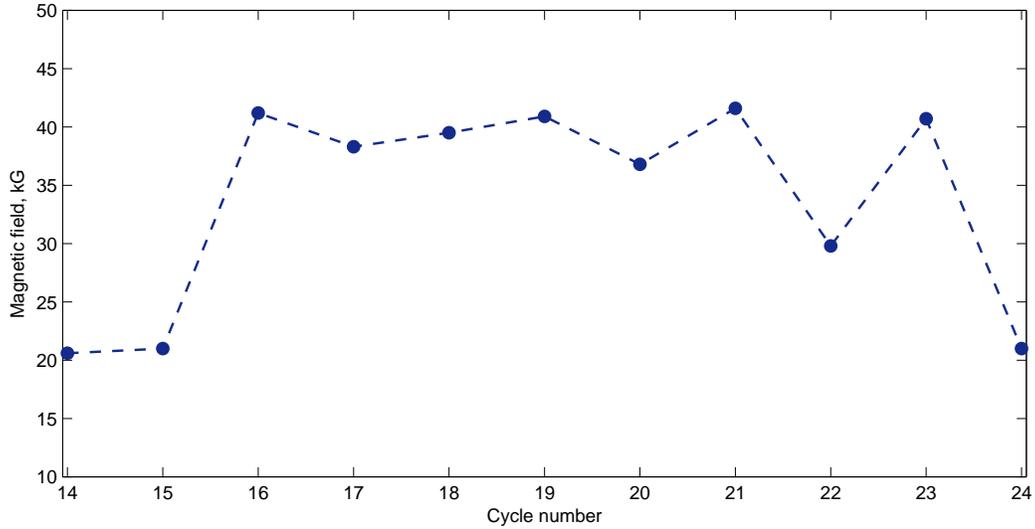}
\caption{Strength of dynamo magnetic field during last 11 sunspot cycles (cycles 14-24) estimated from observed Rieger-type periodicity.}
\end{figure}

The two solutions of Eq.~(20) are
\begin{equation}
\label{disp_f} \omega_f={-m \Omega_0}{{1+s_2 + \sqrt{(1+s_2)^2+{{B^2_0} \over {4 \pi \rho \Omega^2_0 R^2_0 }}{n(n+1)}}}\over{n(n+1)}}
\end{equation}
for the fast magnetic Rossby waves and
\begin{equation}
\label{disp_s} \omega_s
={-m \Omega_0}{{1+s_2 - \sqrt{(1+s_2)^2+{{B^2_0} \over {4 \pi \rho \Omega^2_0 R^2_0 }}{n(n+1)}}}\over{n(n+1)}},
\end{equation}
for the slow magnetic Rossby waves, respectively. The slow magnetic Rossby waves have very long periods (hundreds of years) for a weak magnetic field strength of $B_0=1\;$kG \citep{zaqarashvili2015}, but these periods approach to 1 yr for stronger magnetic field strengths of $B_0=100\;$kG. These waves are not relevant for the present paper, but they may be important to explain the recently found $\sim$ 1 yr periodicity \citep{McIntosh2015}. On the other hand, the fast magnetic Rossby waves have periods in the range of the Rieger periodicity. Eq.~(16) shows that the actual strength of the toroidal magnetic field varies along the latitude and it has a maximal value
\begin{equation}
\label{Bmax} B_{max}={{B_0}\over {2}},
\end{equation}
at a latitude of $45$ degrees. Therefore, the actual strength of the toroidal component at mid latitudes is half of the magnetic field strength, $B_0$, that appears in the dispersion relation.

Figure~3 shows the dependence of the fast magnetic Rossby wave period on the toroidal magnetic field strength, $B_{max}$, using Eq.~(21) for two spherical harmonics with $m=1$ and $n=3,4$. It is clearly seen that the period of the fast magnetic Rossby waves with $m=1$ and $n=4$ is in the range of the Rieger periodicity for $30-50\;$kG. The harmonic with $m=1$ and $n=3$ provides shorter periods of $<140\;$days for any smaller value of the magnetic field strength (not shown here). Therefore, the variation of the toroidal magnetic field strength from cycle to cycle may lead to different observed periods for the $m=1$ and $n=4$ harmonic of the fast magnetic Rossby waves.

\section{Estimation of dynamo magnetic field strength near the base of convection zone during cycles 14-24}

In this section, we use the dispersion relation of the fast magnetic Rossby waves, i.e.\ Eq.~(21), to estimate the strength of the dynamo magnetic field during cycles 14-24.

The location of the region in the solar interior, where the dynamo magnetic field is generated,  is an open question. Most of the dynamo models suggest that the amplification of the magnetic field occurs in the tachocline \citep{Parker1993,Ferriz-Mas1994}. However, recent 3D numerical simulations including convection, rotation and the magnetic field, showed that a stable toroidal magnetic field can be formed in the convection zone without considering the tachocline \citep{Brown2010,Brown2011,Ghizaru2010,Nelson2011}. The interested reader may look into \citet{Charbonneau2013}, who summarized the remaining  problem and the results obtained by the recent numerical simulations. Three from these simulations \citep{Brown2010,Brown2011,Nelson2011} were performed for the young Sun with a much faster rotation (3-5 times faster than the current rotation). \citet{Ghizaru2010}, however, used current solar parameters and found that the amplification of the magnetic field and periodic reversals (with a period of about 30 years, which is three times longer than current solar cycles) may take place near the bottom of the convection zone, but without considering a stable thin layer like the tachocline. Therefore, in the following we define the dynamo layer as the region of dynamo action, which may correspond to the tachocline, or it may lie
somewhere in the convection zone. The dynamo magnetic field strength obtained by \citet{Ghizaru2010} was much smaller (about $2.5\;$kG) than that is usually suggested in the tachocline (viz.\ about $10-100\;$kG). Therefore, the magnetic field strength may appear as a key to find the location of the dynamo layer. We will use our formalism to estimate the magnetic field strength in the dynamo layer at the bottom of convection zone. Then the field strength may put some constraints on the dynamo models with and without tachocline.

According to Eq.~(21), the period of the spherical harmonics with fixed $m$ and $n$ depends on the amplitude of the magnetic field strength, $B_0$, the distance from the center to the dynamo layer, $R_0$, the equatorial angular velocity, $\Omega_0$, and the differential rotation rate, $s_2$. However, these parameters vary from cycle to cycle, except for $R_0$. Note that the distance of the dynamo layer from the solar center in \citet{Ghizaru2010} almost coincides to the location of tachocline, therefore $R_0$ has a similar value for models with and without tachocline. The surface values of $\Omega_0$ and $s_2$ can be measured from observations (there could be slight difference between the values at the surface and the bottom of convection zone, but it will not significantly affect the estimation), but the strength of the dynamo magnetic field is not known. Therefore, using the observed Rieger period, the equatorial angular velocity and the differential rotation rate, one can estimate the amplitude of the magnetic field strength, $B_0$ and hence $B_{max}$, for each individual cycle.

Several authors studied the long-term variation of the differential rotation parameters and estimated the values of the equatorial angular velocity and the differential rotation rate through the different solar cycles \citep{Javaraiah2005,Brajsa2006,Zhang2011}. We here use the differential rotation parameters obtained by \citet{Javaraiah2005} for sunspot cycles 12-23 (see table~1). For the current cycle 24, we use the averaged values of the equatorial angular velocity and the differential rotation rate over the previous 10 cycles. Based on these parameters, we calculate the dynamo magnetic field strength for cycles 14-24 using Eq.~(21) for the harmonic with $m=1$ and $n=4$ (the harmonic with $m=1$ and $n=3$ does not yield the observed periodicity for any value of the magnetic field). The resulting estimated field strength is displayed on Table~1 (last column) and plotted on Figure~4. The magnetic field strength during cycles 14, 15, 24 are estimated to be $\sim 20\;$kG, while the cycles 16-23 show a mean field strength of $35-40\;$kG. The magnetic field strength is generally correlated with the sunspot cycle amplitude as expected from physical grounds. The dynamo magnetic field in cycle 22 is a bit smaller ($30\;$kG), while the cycle was strong enough. This discrepancy is probably caused by observed long period in the GRO data (181 days). On the other hand, the ROB data shows a stronger peak at 157 days, yielding a magnetic field strength of $\sim 40\;$kG, which seems more appropriate for  cycle 22. This cycle shows that the method of magnetic field strength estimation needs to be applied with a certain caution and the errors could be quite high. Therefore, a careful statistical analysis with several different sources and data is required to improve the precision of the estimation. Additionally, Eqs.~(9)-(10) are solved analytically for lower latitudes and small differential rotation. Therefore, the resulting estimated magnetic field strength is a rather rough approximation. Direct numerical simulation may increase the accuracy of the magnetic field estimation.

\begin{table}[h!]
\centering
 \begin{tabular}{||c c c c c c c  ||}

  \hline
 cycle  & period  & period  & $s_2$ & $\Omega_0$ &  $B_{max}, kG $ \\
 number &(days, GRO) & (days, ROB) & & (s$^{-1}$)& GRO (ROB) & \\ [0.5ex]
 \hline\hline
 14 & 195 & 197 & 0.15 & 2.927 10$^{-6}$ & 20.6 (19.4) \\
 \hline
 15 & 190 & 185 & 0.18 & 2.931 10$^{-6}$ & 21 (24.1) \\
 \hline
 16 & 158 & 160 & 0.17 & 2.929 10$^{-6}$ & 41.2 (40) \\
 \hline
 17 & 160 & 161 & 0.20 & 2.933 10$^{-6}$ & 38.3 (37.6) \\
 \hline
 18 & 160 & 158 & 0.18 & 2.92 10$^{-6}$  & 39.5 (40.8) \\
 \hline
 19 & 157 & 167 & 0.19 & 2.925 10$^{-6}$ & 40.9 (34.6) \\
 \hline
 20 & 166 & 160 & 0.16 & 2.927 10$^{-6}$ & 36.8 (40.5) \\
 \hline
 21 & 160 & 155 & 0.14 & 2.925 10$^{-6}$ & 41.6 (44.7) \\
 \hline
 22 & 181 & 157 & 0.14 & 2.909 10$^{-6}$ & 29.8 (43.8) \\
 \hline
 23 & 159 & 150 & 0.17 & 2.92 10$^{-6}$ & 40.7 (46.6) \\
 \hline
 24 & 192 & 197 & 0.168 & 2.92 10$^{-6}$ & 21 (17.8) \\ [1ex]
 \hline
\end{tabular}
\caption{Estimated Rieger periods from GRO and ROB data, differential rotation parameters $\Omega_0$ and $s_2$ from \citet{Javaraiah2005}, and estimated dynamo magnetic field strength for solar cycles 14-24.}
\label{table:1}
\end{table}

\section{Discussion}

The Rieger-type periodicity of 154-160 days observed in the solar magnetic activity variation is one of the unsolved problems of solar physics. It is clear that the periodicity is not a permanent feature: it appears only from time to time, persists during several years and then disappears again. It was shown in previous studies that the periodicity usually appears during maxima of the solar magnetic cycles \citep{Lean1990,Oliver1998,zaqarashvili2010a}. However, not all cycles show evidence of the periodicity. Several different mechanisms have been suggested, but none of these could explain why the periodicity is seen in one cycle, but not in another one. The only possibility is to suppose that the periodicity strongly depends on the solar cycle strength and, hence, on the dynamo magnetic field. This hypothesis is strengthened by the fact that the periodicity is also seen in emerging magnetic flux, which clearly indicates to its connection to deeper layers. \citet{zaqarashvili2010a} suggested that the Rieger periodicity can be explained by magnetic Rossby waves in the solar tachocline. The magnetic Rossby waves are related to the HD Rossby waves but affected by the toroidal magnetic field. Therefore, the periods of spherical harmonics of the waves depend on the magnetic field strength \citep{zaqarashvili2007,zaqarashvili2009,zaqarashvili2011}. Consequently, the variation of the mean dynamo field (which actually defines the strength of the solar cycles) from cycle to cycle may lead to a significant variation of the observed Rieger-type periodicity.

In this paper, we have studied long-term sunspot data sets from the Greenwich Royal Observatory and the Royal Observatory of Belgium for cycles 14-24, using a Morlet wavelet method. Our analysis showed that the Rieger-type periodicity appears in all cycles, but that the periods are varying from 155 to 200 days from cycle to cycle. Figure~1 clearly showed that the periodicity correlates with the solar magnetic activity: shorter periods generally appear during stronger cycles. This means that the dilemma of the disappearance of the periodicity in some cycles can be easily solved: the periodicity does not disappear but it changes. For example, \citet{Carbonell1992} analysed the cycles 12-21 and showed that the periodicity of 155   days was absent in the cycles 12-15, but it was present in cycles 16-21. A careful look at Figure~1 of the present paper shows that the periodicity during cycles 16-21 is about 155-160 days, in complete agreement with \citet{Carbonell1992}. On the other hand, the periodicity during cycles 14-15 is 185-190 days in our analysis, while it was absent in the analysis of \citet{Carbonell1992} as these authors only looked for a periodicity of 155-160 days. We conclude that the cycle strength defines the value of Rieger-type periodicity. This also means that the Rieger-type period during individual cycles can give information about the strength of the cycle. As the cycle strength is probably connected with the dynamo magnetic field, the Rieger periodicity may then provide a tool to estimate the field strength in individual cycles.

Different dynamo models predict different locations of the layer where the solar magnetic field is amplified. The traditional approach is that the dynamo layer is located below the convection zone, where the transition between the differentially rotating convection zone and the uniformly rotating radiative envelope takes place. This thin stable transition layer, called the tachocline \citep{Spiegel1992}, permits the storage of a strong toroidal magnetic field. This magnetic field then becomes unstable due to magnetic buoyancy and as a result magnetic flux rises upwards in the form of magnetic flux tubes, which appear as sunspots at the surface \citep{Parker1993,Ferriz-Mas1994,Fan2009,Charbonneau2010}. The magnetic field strength can be estimated from simulations of rising magnetic flux tubes, which yields values of $10-100\;$kG \citep{Schussler1994,Rempel2006}. On the other hand, recent 3D simulations with different numerical methods suggest that the amplification of a stable toroidal magnetic field may happen in the convection zone even without the existence of a tachocline \citep{Brown2010,Brown2011,Ghizaru2010,Nelson2011}. These simulations mostly concern the young Sun with faster rotation, but they caused some suspicion regarding the significance of the tachocline in the formation of the current solar cycles. In particular, \citet{Ghizaru2010} used the current solar parameters and could reproduce the toroidal magnetic field with periodic reversals without the need of a tachocline. The maximum magnetic field was achieved near the base of the convection zone with a strength of $2.5\;$kG (though the field strength may depend on parameters of solar convection during simulations). Hence, the dynamo magnetic field models with a tachocline predict stronger toroidal fields than those without a tachocline. Since the Rieger periodicity correlates to the strength of dynamo magnetic field, it may place some constraints on these dynamo models. However, one still needs a physical mechanism that is responsible for causing the periodicity.

In the present paper, we suggest that the Rieger periodicity is caused by magnetic Rossby waves within the dynamo layer. The Rossby waves may occur on a 2D spherical surface and, hence, may work in any dynamo model with or without a tachocline. We derived the dispersion relation of the magnetic Rossby waves modified by the existing latitudinal differential rotation (Eq.~20). This equation is similar to Eq.~(5) in \citet{zaqarashvili2009}, but the additional constant $s_2$ appears as a result of taking into account the differential rotation. When $s_2=0$, then Eq.~(20) reduces to Eq.~(5) of \citet{zaqarashvili2009}. Figure~3 shows that the period of the spherical harmonic of the fast magnetic Rossby waves with $m=1$ and $n=4$ gives a periodicity of 150-190 days for a toroidal magnetic field strength of $30-50\;$kG. Therefore, this harmonic can be responsible for the observed Rieger periodicity in different solar cycles. It was shown that the joint action of differential rotation and the toroidal magnetic field may lead to a global MHD instability in the tachocline \citep{Dikpati1999,Dikpati2003,Dikpati2005}. This is probably the same instability found by \citet{zaqarashvili2010a,zaqarashvili2010b} for magnetic Rossby waves through the method of Legendre polynomial expansion. \citet{Dikpati1999,Dikpati2003,Dikpati2005} used an inertial frame in the analysis, while \citet{zaqarashvili2010a,zaqarashvili2010b} used a rotating frame. It is well known that the Rossby wave spectrum generally appears in the frame co-rotating with a sphere. Therefore, we suggest that global MHD instabilities found by \citet{Dikpati1999,Dikpati2003,Dikpati2005} will show the magnetic Rossby wave spectrum if a rotating frame would be used instead of an inertial frame.

Using the dispersion relation (Eq.~21) of the $m=1$ and $n=4$ spherical harmonic and the observed Rieger-type periods from the RGO and the ROB sunspot data for cycles 14-24, we then estimated the toroidal magnetic field strength below the solar convection zone during these cycles. Figure~4 shows the resulting magnetic field strength. In most cycles (16-23) the estimated magnetic field strength is $\sim 40\;$kG. However, in very weak cycles (such as cycles 14, 15 and 24) the estimated field strength is reduced to $20\;$kG. The toroidal magnetic field strength is very important to understand the solar dynamo. Our results show that the estimated strength of the toroidal magnetic field corresponds to the dynamo models with a tachocline rather than to those without a tachocline. This is because the estimated strength of $\geq 20\;$kG is too high for the dynamo models where the tachocline is absent \citep{Ghizaru2010}. However, the estimated magnetic field strength is a rather rough approximation as the dispersion relation is obtained analytically for lower latitudes and a small differential rotation. Other more complex methods, for example, a Legendre polynomial expansion, may yield a smaller estimate for the strength of the toroidal field \citep{zaqarashvili2010a,zaqarashvili2010b}. On the other hand, even dynamo models with a tachocline predict controversial values of the magnetic field strength. As a matter of fact, early estimations of the toroidal field strength inferred from simulations of rising magnetic flux tubes \citep{Schussler1994} showed values around $100\;$kG, which is orders of magnitude larger than the equipartition field strength estimated from the convection. On the other hand, dynamic flux-transport dynamo models that include the feedback of the induced magnetic field on differential rotation and meridional flow estimated the saturated field strength at the base of the convection zone as $12-15\;$kG \citep{Rempel2006}. Yet, recent 3D numerical simulations of rising flux tubes through a convection zone, which include the coupling between flux tube and surrounding convective flow field, found best agreement with observed active region properties for the field strength of $\sim 40-50\;$kG \citep{Weber2011,Weber2013}. The results of these 3D simulations are in very good agreement with our estimations for strong cycles 16-23. On the other hand, the weaker cycles (14, 15 and 24) with the magnetic field strength of $20\;$kG may lead to slightly different properties of active regions (tilt, latitudes, etc.). The idea can be tested by comparison of observations during the weak cycles and 3D simulations using a field strength of $20\;$kG. This can be done for the current cycle 24 due to the better possibility to trace the properties of active regions.

\section{Conclusion}

A wavelet analysis of the Greenwich Royal Observatory (USAF/NOAA) and the Royal Observatory of Belgium (WDC-SILSO) sunspot data for solar cycles 14-24 showed the existence of a Rieger-type periodicity in all cycles. The sunspot cycles 14, 15, 24 and 16-23 displayed a periodicity of 185-195 and 155-165 days, respectively. It is found that the periodicity correlates with the cycle strength: stronger cycles show shorter periods. The periodicity is explained in terms of magnetic Rossby waves near the base of convection zone. The periods of the spherical harmonic of fast magnetic Rossby waves with $m=1$ and $n=4$, where $m$ ($n$) denotes toroidal (poloidal) wave number, perfectly fit with the observed periodicity for the cycle-dependent toroidal magnetic field. The observed Rieger-type periodicity is then used to estimate the dynamo magnetic field strength during cycles 14-24. The estimations predict the magnetic field strengths of $\sim 20\;$kG and $\sim\;$40 kG for cycles 14, 15, 24 and 16-23, respectively, in the dynamo layer, which may correspond to the tachocline or may lie somewhere in the convection zone. The estimated field strength corresponds to the dynamo models with tachocline rather than to those without tachocline, but more sophisticated methods may change this situation. The estimated strength is in good agreement with recent 3D numerical simulations of rising flux tubes through a convection zone, which include the coupling between flux tube and surrounding convective flow field \citep{Weber2011,Weber2013}. A careful analysis of the different data sets using different statistical methods may improve the values of Rieger periods in different cycles, while numerical simulations may lead to more precise estimations of dynamo magnetic field strength.

{\bf Acknowledgements} The work was supported by the Austrian "Fonds zur F\"{o}rderung der Wissenschaftlichen Forschung" (FWF) under projects P26181-N27 and P25640-N27 and by FP7-PEOPLE-2010-IRSES-269299 project- SOLSPANET. Source of yearly mean sunspot numbers: WDC-SILSO, Royal Observatory of Belgium, Brussels. We thank the referee for his/her very constructive comments.

\appendix

\end{document}